\documentclass[12pt]{iopart}

\usepackage[dvips]{graphicx}
\usepackage{array}
\usepackage{amssymb}
\usepackage{hhline}
\usepackage{longtable}
\usepackage{dcolumn}
\usepackage{bm}
\usepackage{subfigure}
\usepackage[hyperindex,CJKbookmarks,dvipdfm,colorlinks,citecolor=blue,linkcolor=blue]{hyperref}
\begin{document}

\title{Average distance in a hierarchical scale-free network: an exact solution}

\author{Zhongzhi Zhang$^1$${}^,$$^2$, Yuan Lin$^1$${}^,$$^2$, Shuyang Gao$^1$${}^,$$^2$, Shuigeng
Zhou$^1$${}^,$$^2$, and Jihong Guan$^3$}

\address{$^1$ School of Computer Science, Fudan University,
Shanghai 200433, China}
\address{$^2$ Shanghai Key Lab of Intelligent Information
Processing, Fudan University, Shanghai 200433, China}
\address{$^3$ Department of Computer Science and Technology,
Tongji University, 4800 Cao'an Road, Shanghai 201804, China}

\eads{zhangzz@fudan.edu.cn,sgzhou@fudan.edu.cn,jhguan@tongji.edu.cn}

\begin{abstract}
Various real systems simultaneously exhibit scale-free and
hierarchical structure. In this paper, we study analytically average
distance in a deterministic scale-free network with hierarchical
organization. Using a recursive method based on the network
construction, we determine explicitly the average distance,
obtaining an exact expression for it, which is confirmed by
extensive numerical calculations. The obtained rigorous solution
shows that the average distance grows logarithmically with the
network order (number of nodes in the network). We exhibit the
similarity and dissimilarity in average distance between the network
under consideration and some previously studied networks, including
random networks and other deterministic networks. On the basis of
the comparison, we argue that the logarithmic scaling of average
distance with network order could be a generic feature of
deterministic scale-free networks.
\end{abstract}


\maketitle


\section{Introduction}

In the last decade, a lot of authors in different scientific
communities have made a concerted effort toward unveiling and
understanding the generic properties of complex networked systems in
nature and society~\cite{AlBa02,DoMe02,Ne03,BoLaMoChHw06}. One of
the most important discoveries is that despite network diversity,
most real-life networks exhibit striking small-world
behavior~\cite{WaSt98}, that is to say, their average distance
scales logarithmically with network order (number of nodes in a
network), or slowly. Average distance is a fundamental measurement
characterizing a complex network, which is relevant to many other
structural features of the network, including degree
distribution~\cite{ChLu02,CoHa03}, centrality~\cite{DoMeOl06},
fractality~\cite{SoHaMa06, ZhZhZo07,ZhZhChGu08}, and so on. In
addition, average distance has a strong effect on various dynamics
running on networks, such as disease spreading~\cite{WaSt98}, random
walks~\cite{CoBeTeVoKl07}, synchronization~\cite{NiMoLaHo03},
amongst others. In view of its significance and usefulness, average
distance has received considerable
attention~\cite{DoMeSa03,Lolo03,FrFrHo04,HoSiFrFrSu05,ZhZhChYiGu08,ZhChFaZhZhGu09,FeVaPo09}.

Apart from the small-world feature, a variety of real networks,
particularly biological and social networks, also share two
remarkable properties: scale-free behavior~\cite{BaAl99} and
hierarchical structure~\cite{RaSoMoOlBa02,ClMoNe08}. To mimic
simultaneously the two prominent characteristics, Barab\'asi, Ravasz
and Vicsek proposed a deterministic model~\cite{BaRaVi01}, hereafter
called BRV model, which is the progenitor of deterministic models
for complex networks and has led to an increasing number of
theoretical investigations on deterministic networks that are an
interesting class of networks and have been proved to be a useful
tool~\cite{DoGoMe02,JuKiKa02,
RaBa03,No03,NoRi04a,HiBe06,RoHaAv07,Hi07,ZhZhChGu07,ZhZhFaGuZh07,CoMi09,ZhGuDiChZh09}.
Many structural and dynamical properties of the BRV model have been
studied in much detail, including degree
distribution~\cite{BaRaVi01}, spectra of adjacency
matrix~\cite{IgYa05}, random walks~\cite{AgBu09}, to name but a few.
However, in spite of its importance, the exact knowledge of average
distance for the BRV model remains not well understood.

To fill this gap, in this paper, we study the average distance in
the BRV model, the deterministic nature of which makes it possible
to investigate analytically the average distance. Based on the
recursive relations derived from the self-similar structure of the
BRV model, we obtain the closed-form solution for the average
distance. The obtained rigorous result shows that the average
distance behaves logarithmically with the network order. This
logarithmic scaling has also been previously reported for many other
deterministic scale-free networks. We thus conjecture that the
logarithmic scaling characterizes the behavior of average distance
for deterministic scale-free networks.

\section{The hierarchical scale-free network}

\begin{figure}
\begin{center}
\includegraphics[width=0.8\linewidth,trim=100 280 100 220]{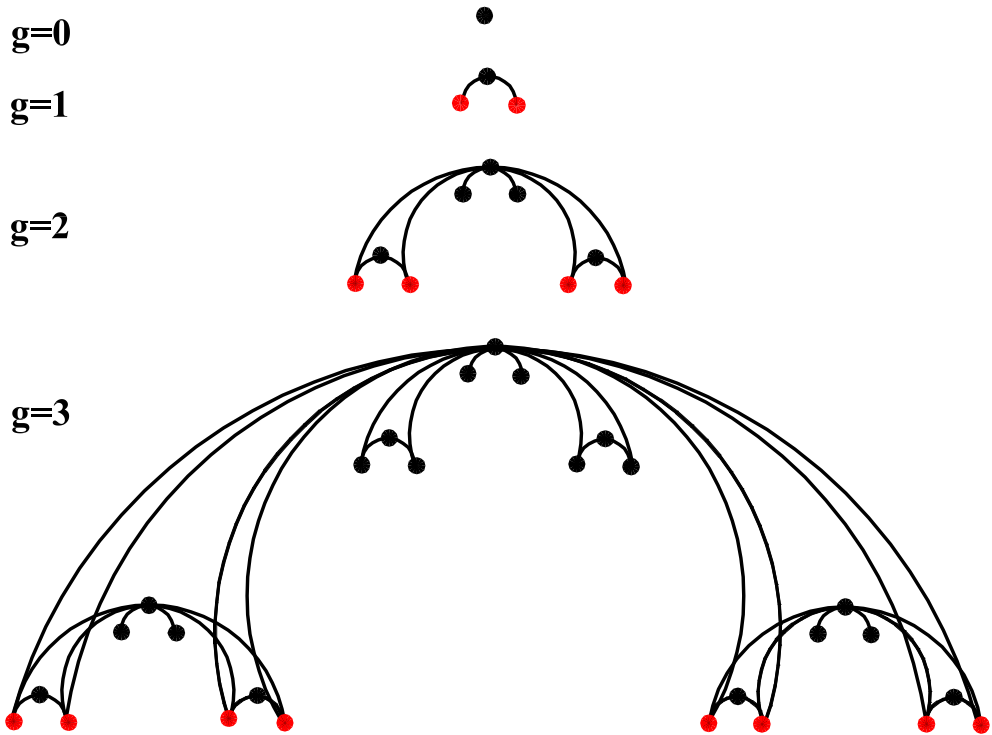}
\end{center}
\caption[kurzform]{\label{network} (Color online.) The iterative
construction process of the BRV hierarchical scale-free network,
showing only the first few iterations. The bottoms nodes are
depicted in red.}
\end{figure}

Let us first introduce the BRV model, a hierarchical scale-free
network that is constructed in an iterative way~\cite{BaRaVi01}. We
denote by $H_{g}$ the BRV network model after $g$ ($g\geq 0$)
iterations (number of generations). Initially ($g=0$), the network
$H_{0}$ consists of a single root node labeled as $i=1$. At the
first generation ($g=1$), two new nodes $i=2,3$ are added to the
system and connected to the root node. Thus, we get $H_{1}$, where
node 1 is called hub node, the nodes $i=2,3$ are named bottom nodes
forming a set represented as $\mathbb{B}_1=\{2,3 \}$. At generation
2 (i.e., $g=2$), we generate two copies of $H_{1}$ and connect the
bottom nodes of each replica to the hub of the original $H_{1}$. The
hub of the original $H_{1}$ and the four bottom nodes in the
replicas become the hub and bottom nodes of $H_{2}$, respectively.
The set of the bottom nodes belonging to $H_{2}$ is denoted as
$\mathbb{B}_2$. Suppose one has $H_{g-1}$; the next generation
network $H_{g}$ can be obtained from $H_{g-1}$ by adding two copies
of $H_{g-1}$ with their bottom nodes being linked to the hub of the
original $H_{g-1}$. In $H_{g}$, its hub is the hub of the original
$H_{g-1}$, its bottom nodes are composed of all the bottom nodes of
both copies of $H_{g-1}$, and all the bottom nodes make the set
$\mathbb{B}_g$. Repeat indefinitely the replication and connection
steps, we obtain the hierarchical scale-free network.
Figure~\ref{network} illustrates the construction process of the
network for the first three iterations.

Some properties of the BRV model have been investigated in
detail~\cite{IgYa05}. Let $N_g$ be the number of nodes in $H_{g}$,
the network of $g$th generation. By construction, at each new
iteration, the number of network nodes increases by a factor of
three, which together with the initial condition $N_{0}=1$ leads to
$N_g=3^{g}$. In $H_{g}$, the degree of all nodes and the number of
nodes having the same degree can be determined
exactly~\cite{IgYa05}. For example, the degree of the hub node is
$K_h(g)=2(2^g-1)$, and the degree of bottom nodes is $K_b(g)=g$.
Again for instance, the cardinality, defined as the number of nodes
in a set, of the set for the bottom nodes is $|\mathbb{B}_g|=2^g$.
The network is a sparse one with the mean degree averaged over all
nodes being $\langle k \rangle_g = 4\left
[1-\left(\frac{2}{3}\right)^g\right]$ which is approximately equal
to $4$ in the limit of infinite $g$.

The BRV model presents some typical properties of real-life
systems~\cite{BaRaVi01,IgYa05}. It is scale-free with the degree
distribution exponent $\gamma=1+\frac{\ln 3}{\ln 2} $. In
particular, the network has a crucial feature characterized by an
obvious hierarchical structure that has also been observed in many
real networks, e.g., metabolic networks~\cite{RaSoMoOlBa02,RaBa03}.
All these characteristics are not shared by other previous models.
The peculiar structural characteristics make the network unique
within the category of scale-free networks. It is the precursor,
probably the first model for hierarchical scale-free networks.
However, in spite of its importance, the rigorous knowledge of the
average distance is still missing; its exact determination is the
primary topic of this paper.

\section{Closed-form solution to average distance}

After introducing the hierarchical scale-free network, we now derive
analytically the average distance. We represent all the shortest
path lengths of network $H_{g}$ as a matrix in which the entry
$d_{ij}(g)$ is the distance between nodes $i$ and $j$ that is the
length of a shortest path joining $i$ and $j$. A measure of the
typical separation between two nodes in $H_{g}$ is given by the
average distance $d_{g}$ defined as the mean of distances over all
pairs of nodes:
\begin{equation}\label{apl01}
d_{g}  = \frac{D_g}{N_g(N_g-1)/2}\,,
\end{equation}
where
\begin{equation}\label{total01}
D_g = \sum_{i \in H_{g},\, j \in H_{g},\, i \neq j} d_{ij}(g)
\end{equation}
denotes the sum of the distances between two nodes over all couples.
Notice that in Eq.~(\ref{total01}), for a pair of nodes $i$ and $j$
($i \neq j$), we only count $d_{ij}(g)$ or $d_{ji}(g)$, not both.

\begin{figure}
\begin{center}
\includegraphics[width=0.8\linewidth,trim=80 180 260 130]{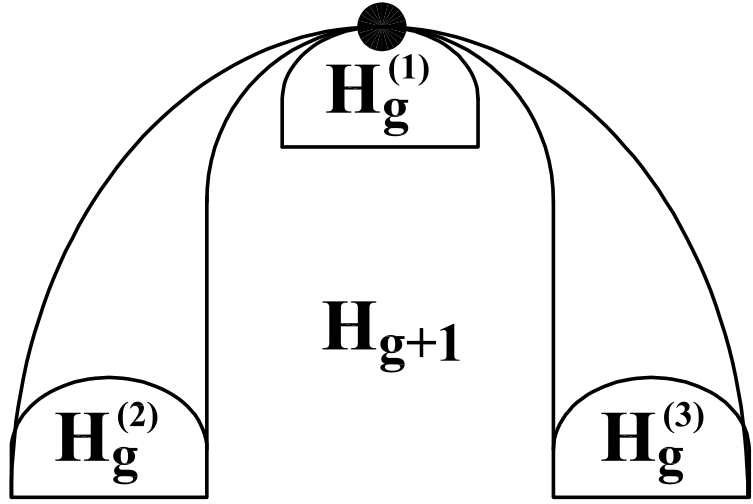}
\end{center}
\caption[kurzform]{\label{labeling} Schematic illustration of the
means of construction of the hierarchical scale-free network.
$H_{g+1}$ is obtained by joining three replicas of $H_{g}$ denoted
as
    $H_{g}^{(\varphi)}$ $(\varphi=1,2,3)$, which are
    connected to one another at the hub node of $H_{g}^{(1)}$. The black node in the figure is the hub denoted by $X$ (not labeled).}
\end{figure}

We continue by exhibiting the procedure of determining the total
distance and present the recurrence formula, which allows us to
obtain $D_{g+1}$ of the $g+1$ generation from $D_{g}$ of the $g$
generation. The hierarchical network $H_{g}$ has a self-similar
structure that allows one to calculate $D_{g}$ analytically.
According to the construction, see figure~\ref{labeling}, network
$H_{g+1}$ is obtained by joining three copies of $H_{g}$ that are
labeled as $H_{g}^{(1)}$, $H_{g}^{(2)}$, and $H_{g}^{(3)}$. Using
this self-similar property, the total distance $D_{g+1}$ satisfies
the recursion relation
\begin{equation}\label{total02}
  D_{g+1} = 3\, D_g + \Delta_g,
\end{equation}
where $\Delta_g$ is the sum over all shortest path length whose
endpoints are not in the same $H_{g}^{(\varphi)}$ branch. The paths
that contribute to $\Delta_g$ must all go through the hub node $X$,
where the three copies of $H_{g}$ are connected. Hence, to determine
$D_g$, all that is left is to calculate $\Delta_g$. The analytic
expression for $\Delta_g$, referred to as the crossing path length,
can be derived as below.

Let $\Delta_g^{(\alpha,\beta)}$ be the sum of the lengths of all
shortest paths whose endpoints are in $H_{g}^{(\alpha)}$ and
$H_{g}^{(\beta)}$, respectively. Then the total sum $\Delta_g$ is
given by
\begin{equation}\label{cross01}
\Delta_g =\Delta_g^{(1,2)} + \Delta_g^{(1,3)} + \Delta_g^{(2,3)}\,.
\end{equation}
By symmetry, $\Delta_g^{(1,2)} = \Delta_g^{(1,3)}$, so
\begin{equation}\label{cross02}
\Delta_g =2\,\Delta_g^{(1,2)} + \Delta_g^{(2,3)}\,.
\end{equation}
Having $\Delta_g$ in terms of the quantities of $\Delta_g^{(1,2)}$
and $\Delta_g^{(2,3)}$, the next step is to explicitly determine the
two quantities.

To calculate the crossing distance $\Delta_g^{(1,2)}$ and
$\Delta_g^{(2,3)}$, we give the following notation. For an arbitrary
node $v$ in network $H_{g}$, let $f_v(g)$ be the smallest value of
the shortest path length from $v$ to any of the $2^g$ bottom nodes
belonging to $\mathbb{B}_g$, and the sum of $f_v(g)$ for all nodes
in $H_{g}$ is denoted by $F_g$. Analogously, in $H_{g}$ let $h_v(g)$
denote the distance from a node $v$ to the hub node $X$, and let
$M_g$ stand for the total distance between all nodes in $H_{g}$ and
the hub node $X$ in $H_{g}$, including $X$ itself. By definition,
$F_{g+1}$ can be given by the sum
\begin{eqnarray}\label{bottom01}
F_{g+1} &=&\sum_{v\in H_{g}^{(1)}} f_v(g+1)\nonumber +\sum_{v\in
H_{g}^{(2)}} f_v(g+1)+\sum_{v\in H_{g}^{(3)}} f_v(g+1)\nonumber \\
&=& \sum_{v\in H_{g}}[h_v(g)+1]+2\,\sum_{v \in
H_{g}}f_v(g)\nonumber \\
&=&2\,F_{g}+N_g+M_g\,,
\end{eqnarray}
and $M_{g+1}$ can be written recursively as
\begin{eqnarray}\label{hub01}
M_{g+1} &=&\sum_{v\in H_{g}^{(1)}} h_v(g+1)\nonumber +\sum_{v\in
H_{g}^{(2)}} h_v(g+1)+\sum_{v\in H_{g}^{(3)}} h_v(g+1)\nonumber \\
&=& \sum_{v\in H_{g}}h_v(g)+2\,\sum_{v \in
H_{g}}[f_v(g)+1]\nonumber \\
&=&M_g+2\,(F_{g}+N_g)\,.
\end{eqnarray}
Using $N_g=3^g$, and considering $F_1=1$ and $M_1=2$, the
simultaneous equations~(\ref{bottom01}) and~(\ref{hub01}) can be
solved inductively to obtain:
\begin{equation}\label{bottom02}
F_{g} =3^{g-2}(4g-1)
\end{equation}
and
\begin{equation}\label{hub02}
M_{g} =2\times 3^{g-2}(2g+1)\,.
\end{equation}

With above obtained results, we can determine $\Delta_g^{(1,2)}$ and
$\Delta_g^{(2,3)}$, which can be expressed in terms of these
explicitly determined quantities. By definition, $\Delta_g^{(1,2)}$
is given by the sum
\begin{eqnarray}\label{cross03}
\Delta_g^{(1,2)}&=& \sum_{u \in H_{g}^{(1)},\,v \in H_{g}^{(2)}} d_{uv}(g+1)\nonumber \\
 &=&
\sum_{u \in H_{g}^{(1)},\,v \in H_{g}^{(2)}}
\Big[h_u(g)+1+f_v(g)\Big]\nonumber \\
&=&\sum_{v \in H_{g}^{(2)}} \sum_{u \in H_{g}^{(1)}} h_u(g)+\sum_{u
\in H_{g}^{(1)}}\sum_{v \in H_{g}^{(2)}}[1+f_v(g) ]\nonumber
\\&=&N_g\,M_g+(N_g)^2+N_g\,F_g\,.
\end{eqnarray}
Inserting Eqs.~(\ref{bottom02}) and~(\ref{hub02})
into~(\ref{cross03}), we have
\begin{equation}\label{cross04}
\Delta_g^{(1,2)}=9^{g-2}(8g+2)\,.
\end{equation}
Proceeding similarly,
\begin{eqnarray}\label{cross05}
\Delta_g^{(2,3)}&=& \sum_{u \in H_{g}^{(2)},\,v \in H_{g}^{(3)}} d_{uv}(g+1)\nonumber \\
 &=&2\,[(N_g)^2+N_g\,F_g]\nonumber
\\&=&9^{g-2}(8g+8)\,.
\end{eqnarray}
Substituting Eqs.~(\ref{cross04}) and~(\ref{cross05})
into~(\ref{cross02}), we get
\begin{equation}\label{cross06}
\Delta_g=9^{g-2}(24g+12)\,.
\end{equation}
Substituting Eq.~(\ref{cross06}) into~(\ref{total02}) and using the initial value $D_1=4$, 
we can obtain the exact expression for the total distance
\begin{equation}\label{total03}
D_g=4g\times 9^{g-1}\,.
\end{equation}
Then the analytic expression for average distance can be obtained as
\begin{equation}\label{apl02}
 d_g =\frac{8g\times 3^{g-2}}{3^{g}-1}\,.
\end{equation}

\begin{figure}
\begin{center}
\includegraphics[width=.4\linewidth,trim=110 20 110 0]{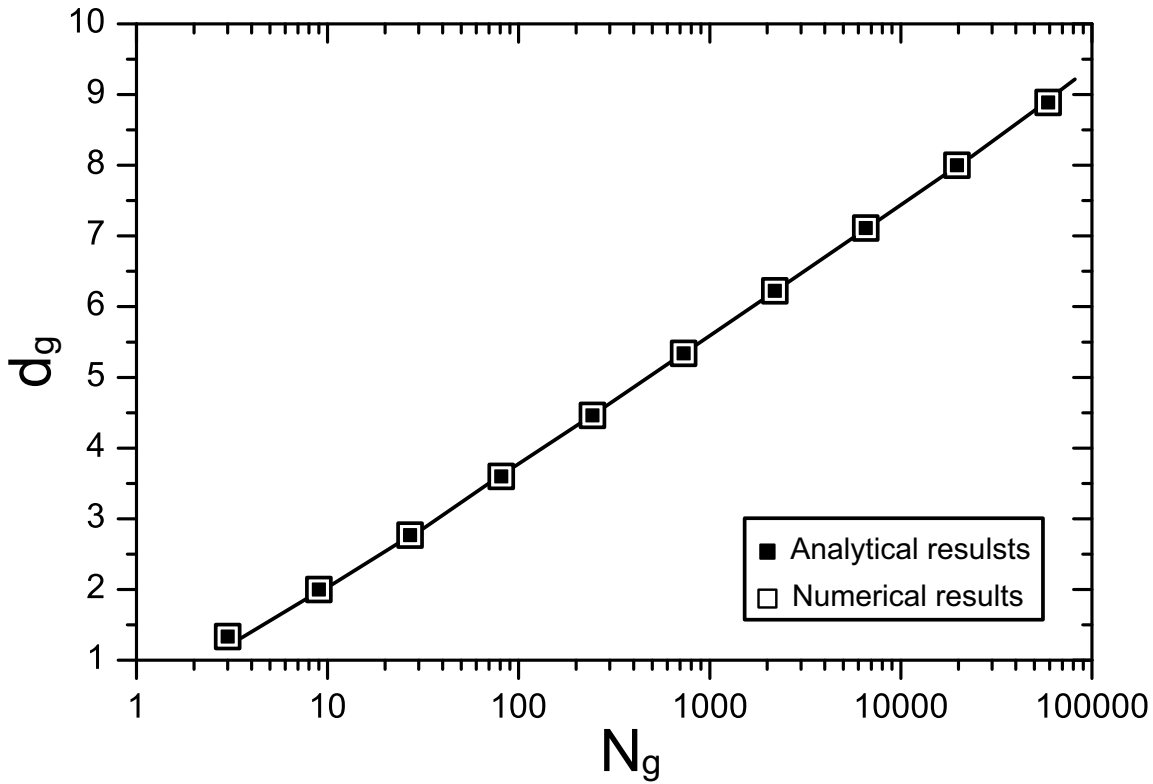}
\end{center}
\caption[kurzform]{\label{AveDis} Average distance $d_{g}$ versus
network order $N_g$ on a semi-logarithmic scale. The solid line is a
guide to the eye.}
\end{figure}

We have checked our rigorous result provided by Eq.~(\ref{apl02})
against numerical calculations for different network order up to
$g=10$ which corresponds to $N_{10}=59\,049$. In all the cases we
obtain a complete agreement between our theoretical formula and the
results of numerical investigation, see figure~\ref{AveDis}.

We continue to express the average distance $d_g$ as a function of
network order $N_g$, in order to obtain the scaling between these
two quantities. Recalling that $N_g=3^g$, we have $g=\log_3N_g$.
Hence Eq.~(\ref{apl02}) can be rewritten as
\begin{equation}\label{apl03}
 d_g =\frac{8N_g\ln N_g}{9\ln 3(N_g-1)}\,.
\end{equation}
In the infinite network order limit, i.e., $N_g\rightarrow \infty$
\begin{equation}\label{apl03}
 d_g =\frac{8}{9\ln 3}\ln N_g\,.
\end{equation}
Thus, for large networks, the average distance grows logarithmically
with increasing order of the network.

This logarithmic scaling is similar to that of other hierarchical
scale-free networks with high clustering coefficient, which was
previously obtained in a quite different way by mapping the system
onto a Potts model in one-dimensional lattices~\cite{No03}. Since
there is no triangle in the studied network, its clustering
coefficient is zero. Our result, together with earlier work, shows
that clustering coefficient has no qualitative effect on the average
distance of deterministic hierarchical scale-free networks, which is
consistent with the phenomenon observed for random scale-free
networks by using numerical simulations~\cite{HoKi02,ZhRoWaZhGu07}.

However, the deterministic network under consideration also exhibits
some different aspects from the conventional (non-hierarchical)
scale-free networks. For example, it has been suggested that for
stochastic scale-free networks with degree distribution exponent
$\gamma <3$ and network order $N$, their average distance $d(N)$
behaves as a double logarithmic scaling with $N$: $d(N)\sim \ln\ln
N$~\cite{CoHa03,ChLu02}, which is in sharp contrast to the
logarithmic scaling obtained for the BRV model addressed here, in
despite of the fact that the latter has a degree distribution
exponent $\gamma=1+\frac{\ln3}{\ln2}$ less than 3. Actually, this
logarithmic scaling of average distance with network order has also
been shown in other deterministic scale-free networks with $\gamma
<3$~\cite{ZhChFaZhZhGu09,DoGoMe02,No03,ZhGuDiChZh09,ZhZhCh07,ZhChZhFaGuZo08}.
Thus, deterministic scale-free networks present an obvious
difference from their stochastic scale-free counterparts in the
aspect of structural property of average distance. We speculate that
the logarithmic scaling for average distance can be used to
establish the universality class for deterministic scale-free
networks. Further studies are necessary to uncover the reasons for
the similarity and dissimilarity between deterministic and random
scale-free networks as regards average distance.

\section{Conclusions}

To conclude, scale-free behavior and hierarchical structure are
ubiquitous in a variety of real-life systems. In this paper, we
studied analytically the average distance of a deterministically
growing scale-free hierarchical network introduced by Barab\'asi,
Ravasz and Vicsek~\cite{BaRaVi01}, which can mimic some real-world
networks to some extent. Based on the particular construction of the
network, we obtained the rigorous solution to the average distance.
We showed that in the infinite limit of network order $N_g$, the
average distance $d_g$ exhibits the scaling law as $d_g \sim \ln
N_g$. We also showed that there are similarity and dissimilarity of
the behaviors of average distance of deterministic and random
scale-free networks. Finally, combining the obtained result and
previous studies, we argued that the logarithmical scaling of
average distance with network order may characterize deterministic
scale-free networks.

\subsection*{Acknowledgment}

This research was supported by the National Basic Research Program
of China under grant No. 2007CB310806, the National Natural Science
Foundation of China under Grant Nos. 60704044, 60873040 and
60873070, Shanghai Leading Academic Discipline Project No. B114, and
the Program for New Century Excellent Talents in University of China
(NCET-06-0376).

\end{document}